# Understanding EFL Students' Idea Generation Strategies for Creative Writing with NLG Tools

David James Woo [a, *], Yanzhi Wang [b], Hengky Susanto [c], and Kai Guo [d]

[a] Precious Blood Secondary School, 338 San Ha Street, Chai Wan, Hong Kong, China

[b] Steinhardt School of Culture, Education and Human Development, New York University, USA

[c] Education University of Hong Kong, Hong Kong, China

[d] Faculty of Education, The University of Hong Kong, Hong Kong, China

[*] Corresponding author

- Postal address: Precious Blood Secondary School, 338 San Ha Street, Chai Wan, Hong Kong, China
- Email address: net_david@pbss.hk
- Phone: +852 2570 4172

## Author Bio

***David James Woo*** is a secondary school teacher. His research interests are in artificial intelligence, natural language processing, digital literacy, and educational technology innovations. ORCID: https://orcid.org/0000-0003-4417-3686

***Yanzhi Wang*** is a graduate student studying Educational Communication and Technology. Her research interests are digital literacy, learning experience design, games for learning/impact, and educational technology innovations.

***Hengky Susanto*** received his BS, MS and PhD degree in computer science from the University of Massachusetts system. He was a post-doctoral research fellow at University of Massachusetts Lowell and Hong Kong University of Science and Technology. He was also senior researcher at Huawei Future Network Theory Lab. Currently, he is a principal researcher in a startup mode research laboratory and a lecturer at Education University of Hong Kong. His research interests include applied AI (computer vision and NLP) to solve complex social problems, smart city, and computer networking (e.g., datacenter network, congestion control, etc.).

***Kai Guo*** is a Ph.D. candidate in the Faculty of Education at the University of Hong Kong. His research focuses on second language writing, technology-enhanced learning, game-based learning, and gamification in education. His recent publications have appeared in international peer-reviewed journals such as *Computers & Education*, *TESOL Quarterly*, and *Assessing Writing*. ORCID: https://orcid.org/0000-0001-9699-7527

**Conflict of Interest Statement**

We have no conflicts of interest to disclose. This research received no specific grant from any funding agency in the public, commercial, or not-for-profit sectors.

# Understanding EFL Students' Idea Generation Strategies for Creative Writing with NLG Tools

## Abstract

Natural language generation (NLG) is a process within artificial intelligence where computer systems produce human-comprehensible language texts from information. English as a foreign language (EFL) students' use of NLG tools might facilitate their idea generation, which is fundamental to creative writing. However, little is known about how EFL students interact with NLG tools to generate ideas. This study explores strategies adopted by EFL students when searching for ideas using NLG tools, evaluating ideas generated by NLG tools and selecting NLG tools for idea generation. Four Hong Kong secondary school students attended workshops where they learned to write stories comprising their own words and words generated by NLG tools. After the workshops, they answered questions to reflect on their writing experience with NLG tools. In a thematic analysis of the written reflections, we found students may have existing ideas when searching for ideas and evaluating ideas with NLG tools. Students showed some aversion to ideas generated by NLG tools and selected NLG tools that generated a greater quantity of ideas. The findings inform our understanding of EFL students' concerns when using NLG tools for idea generation and can inform educators' instruction to implement NLG tools for classroom creative writing.

*Keywords:* Artificial intelligence, natural language generation, idea generation, creative writing, EFL students

## Introduction

Writing is an essential skill by which people can communicate their experiences, feelings and thoughts in a lasting form (Graham & Perin, 2007). For students, writing enhances their reading skills and content area learning, and is crucial for their academic, professional and social success (Graham et al., 2014). Creative writing refers not only to literary art, for example, published fiction, poetry and drama, but importantly, for students, the process by which they produce an original, expressive literary work that aspires to be literary art (Dawson, 2005). Creative writing benefits students not least by enabling students to construct personal knowledge and unique, meaningful insights (Kaufman & Beghetto, 2009). For English as a foreign language (EFL) students, creative writing can increase students' English language proficiency and enable their critical thinking (Dai, 2010). It can have a positive effect on EFL students' writing disposition and achievement (Tok & Kandemir, 2015).

Artificial intelligence (AI) refers to computers that can imitate features of human intelligence such as creative work and language interaction (United Nations Educational, Scientific and Cultural Organization [UNESCO], 2022). Recent advancements in computer network architecture (Vaswani et al., 2017) have greatly improved AI's capacity for natural language generation (NLG), that is, its ability to produce human-comprehensible language texts from information (Gatt & Krahmer, 2018). Thus, NLG tools like ChatGPT can revolutionize the way people interact with written language (Radford et al., 2019). Importantly, unlike automated evaluation systems, the novel capabilities of NLG tools like ChatGPT can only be unlocked when prompted, that is, given a set of instructions (White et al., 2023). Moreover, the content of a prompt greatly impacts this type of NLG tool's ability to generate text so that, for instance, changing the length of a prompt, changing key words in a prompt, or changing the order of words can impact the quality of text generated by the NLG tool. However, a challenge for non-technical users is that crafting an effective prompt that leads to a desired text generated by an NLG tool is not a straightforward process, but one of trial and error (Dang et al., 2022). In this way, a person needs to craft a prompt and then evaluate the quality of the text generated from the NLG tool before using the text in their own writing (Liu et al., 2021).

NLG tools might benefit EFL students' writing by providing more structured assistance to students and reducing their cognitive barriers when writing (Gayed et al., 2022). Moreover, NLG tools might benefit students' creative writing by modeling literary art and by serving as a co-creator with a student in the writing process (Kangasharju et al., 2022). Certainly, prompt-based NLG tools compose an innovative approach to enhance EFL students' creative writing. However, we still have little evidence of how EFL students interact with such tools to support their creative writing. Therefore, the present study seeks to bridge this knowledge gap and inform creative writing instruction by investigating the strategies developed by EFL students when interacting with prompt-based NLG tools to compose short stories.

## Literature Review

### Creative Writing

Creative writing comprises a complex set of cognitive processes (Flowers & Hayes, 1981). One of its fundamental processes is idea generation, which refers to a writer intentionally and strategically searching for ideas, evaluating ideas and translating relevant and important ideas to text until the content for a writing task has been satisfied (Kellogg, 1999). In a classroom writing context, a student's literary work shows the student's competence in idea generation (Rini & Cahyanto, 2020). However, students have struggled with idea generation, not least because children and youths have fewer words, skills and memory resources at their disposal than adults (Lin & Chang, 2020). This struggle is all the more acute for EFL learners who must complete literary work in English language. Besides, students may not receive sufficient writing instruction and feedback from teachers in a classroom context (Butterfuss et al., 2022).

Several studies have investigated pedagogical strategies to enhance students' idea generation, with collaborative writing being a particularly effective approach (Vass, 2002, 2007; Vass et al., 2008). In collaborative writing, two or more writers jointly generate ideas and produce a creative text (Storch, 2011). Interactions during collaborative writing better enable students to generate and formulate their ideas (Zamel, 1982). Experimental studies (Shehadeh, 2011; Khatib & Meihami, 2015) have found collaborative writing interventions have a significant effect on EFL students' ideas. Moonma and Kaweera (2021) found that collaborative writing could improve ideas of novice, intermediate and advanced EFL writers. While collaboration can be an effective means of generating ideas, it can be challenging to find a suitable collaborator with whom students can generate ideas in and out of the classroom. The integration of NLG tools presents a potential solution to this issue. However, to date, schools have rarely deployed NLG tools for classroom writing activities. To realize the potential of NLG technology to transform idea generation and collaboration for EFL students' creative writing will require further research and the expansion of frameworks to analyze interactions and writing products (Li & Zhang, 2023).

**Natural Language Generation Tools**

Studies have demonstrated NLG tools' capability to generate high quality ideas comparable to those from a human. For instance, Brown et al.'s (2020) NLG tool created full-length articles that some people could not distinguish from human-written articles. Similarly, Hitsuwari et al. (2023) found their study's participants could not distinguish between human-made haikus and AI-generated haikus. However, the quality of ideas can depend not least on the type of language model used in an NLG tool and the parameters of that model. In this way, studies have designed, tested and compared language models to improve the quality of ideas for creative writing tasks such as metaphor creation (Gero & Chilton, 2019) and story writing (See et al., 2019; Lee et al., 2022).

The quality of an NLG tool's ideas also depends on how the tool is integrated into a writer's creative writing process. To conceptualize this, Clark et al. (2018) proposed a machine-in-the-loop framework, where the human writer is the central actor and an NLG tool plays a supporting role: at the point where the writer is searching for ideas, the NLG tool is a collaborator that suggests ideas; and the writer has full agency to decide what to do with the NLG tool's ideas, if anything. Importantly, in their experiments, Clark et al. (2018) found that machine-in-the-loop creative writing did not necessarily lead to higher quality stories. Other machine-in-the-loop studies found writers were inspired by NLG tools' ideas and considered NLG tools to be more of an active writer role than a supporting role (Yang et al., 2022). At the same time, individual writers' approaches and motivations could explain differences in machine-in-the-loop writing of stories (Singh et al., 2022). Evidence suggests that students can learn from AI, that is, actively and independently develop strategies to work and to interact with AI to complete a task (Kim et al., 2022). However, to effectively collaborate with prompt-based NLG tools for idea generation, it appears students will need to be taught strategies.

**Writers' Strategies For Using NLG Tools**

Writers require strategies to effectively search for ideas using NLG tools and to evaluate ideas generated from NLG tools. Previous studies have shown some strategies adopted by highly educated, experienced writers. For example, Calderwood et al. (2018) explored how professional novelists wrote with NLG tools. These novelists preferred NLG tools to generate briefer ideas, such as to finish a sentence, than to generate a complete paragraph. Besides, novelists used ideas from the tools to describe story elements, such as scenes and characters and more generally to refine and to elaborate. Yang et al. (2022) found graduate students' strategies for idea generation with AI-generated text to write stories could be classified as those from students who had existing, concrete ideas for their stories and as those who did not. Students with concrete ideas would attempt to find AI-generated ideas that logically fit in their stories; and students without concrete ideas expected less coherence from AI-generated ideas and were more open to new characters, locations and events in AI-generated text.

Although these studies can provide insight into ideas generation strategies with NLG tools, they may not be applicable to EFL students as novice writers. So far, little empirical evidence informs how EFL students might effectively collaborate with NLG tools effectively for idea generation. Thus, the objective of this study is to identify strategies used by EFL students for collaborating and interacting with NLG tools in generating ideas for literary works. With such understanding, writing teachers can better integrate NLG tools into classrooms to enhance their students' creative writing.

**Conceptualizing Student-AI Idea Generation**

Grounded in existing conceptualizations of machine-in-the-loop creative writing (Clark et al., 2018; Yang et al., 2022), we conceptualize an EFL student's idea generation with an NLG tool as a sequence of text-to-text interactions (see Figure 1). To intentionally and strategically search for ideas with an NLG tool, a student prompts, that is, selects an input text for an NLG tool. This prompt may be derived from existing text in the student's literary work. In addition, a student can use the same prompt for one or more NLG tools, or use different prompts for different NLG tools. Based on a prompt, an NLG tool generates output text (Bender et al., 2021), which is the tool's prediction of subsequent words, sentences or paragraphs for the prompt (Liu et al., 2021). The output text represents ideas generated from the NLG tool. These ideas can adhere to a student's prompt. For example, if a student prompts a tool with the first few sentences of a story, the tool can generate a continuation of that story (Hugging Face, n.d.). A student evaluates the ideas generated by the tool and translates any relevant and important ideas, that is, copies and pastes any output text, modifying the literary work as necessary. Otherwise, a student can change the prompt and generate different ideas. A student can use as few or as many ideas from NLG tools as necessary. A student can use NLG tools freely and as many times as necessary to complete a literary work. To achieve effective collaboration with NLG tools and successful idea generation, students need to develop and use strategies for crafting prompts and evaluating ideas from NLG tools.

**Figure 1**

*A conceptual framework of idea generation between a student and NLG tools*

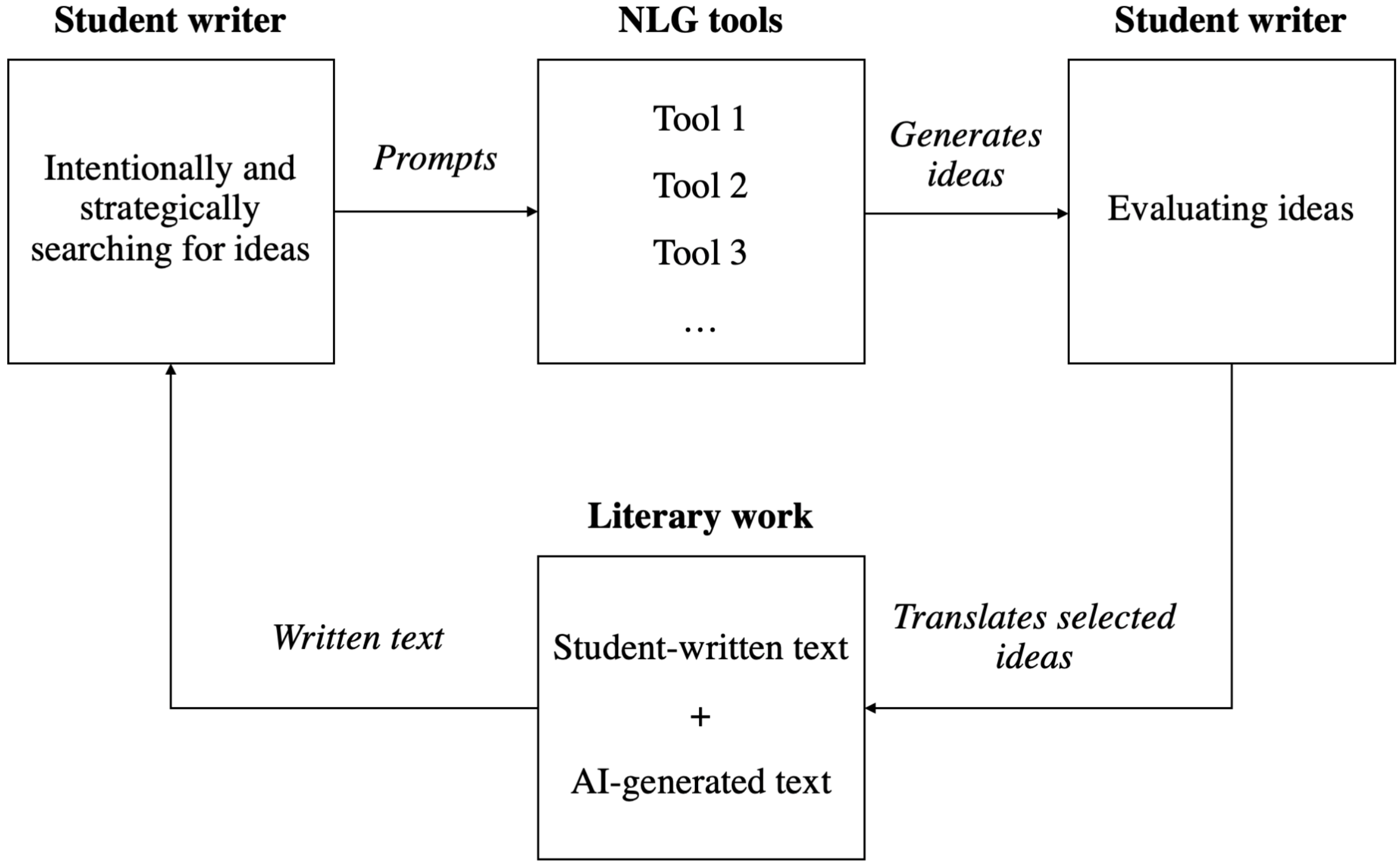

Within this framework, we are interested in investigating not only the strategies by which students intentionally and strategically search for ideas with NLG tools but also the strategies by which students evaluate AI-generated ideas to complete creative literary works. Moreover, we are interested in understanding how students decide with which NLG tools to collaborate for idea generation. To suggest effective strategies, we make intra-student and inter-student comparisons of strategies. The following research questions guide the study.

1) What strategies of intentionally and strategically searching for ideas with NLG tools are taken up by students?
2) What strategies of evaluating ideas from NLG tools are taken up by students?
3) What strategies of selecting NLG tools are taken up by students?

## Methods

### Research Context and Participants

We prosecuted this exploratory study at an all-girls secondary school in Hong Kong. The school primarily delivers instruction to students in Cantonese Chinese language. From February to March, 2022, we organized a voluntary study with the aim for each student participant to write an English language short story of no more than 500 words, using a student's own words and words generated by NLG tools. A student had to write the story on Google Docs and share the doc with the research team. A student had to highlight their own words and distinguish them from AI-generated words (see Figure 2). To facilitate students' writing, we designed seven, voluntary, hour-long workshops from March 1 to April 5, 2022. In these workshops, students learned about the short story text type and its features, AI concepts, prompting approaches for NLG tools and digital writing skills to compose stories on Google Docs; they were given time to plan their short stories, to interact with NLG tools and to compose their stories; and they were given time to ask questions and to reflect. Students did not need to complete their short stories during workshops but submitted their stories on April 7, 2022.

**Figure 2**

*An excerpt from a student's story*

*There is some good in the world which is worth fighting for*

*It's the year 1940,* World War II is raging, and the countries are gripped by fear and anxiety. *You can hear bombs and missals exploding in innocent people houses, young and old both screaming for the sake of there life's. On the other hand, in a peaceful small village on a remote island lives a blind man named Bill, with a blind German Shepherd.*

Bill has no family, and has been living in this village for a very long time. *He met a blind puppy while he was strolling for food one day,* and it has now become his only family. *For Bill and the dog, today was just another peaceful day. They woke up bright early, ate breakfast, went out for a walk, but while they were on the walk* they were interrupted by the sudden appearance of a large force of army, who were talking to each other. *"The sky will get dark soon, let's settle our tent here, and we will destroy them tomorrow." said one of the army. Bill was frightened, and that night Bill informed the entire village that they should evacuate and go to a safer place, but no one believed him,* they thought he was a mad man.

The next day, the army was sent to fight. Everyone thought they would be safe in their house, and the blind man was just joking, but they were attacked from the sky. The army was flying low and they were attacking the village. *After some time, the entire village was flowing in blood, the army were celebrating their victory.*

Note: In this excerpt, a student's words are italicized and AI words are not italicized. In the online version, a student's words are also highlighted in red.

Students were recruited through English language advertisements on campus television and Google Classroom. Four students (N = 4) took part in the study. They are given the pseudonyms Student H, Student M, Student S and Student W in this study. At the time of study, Student H was 16 years old, Student M 12 years old, Student S 15 years old, and Student W 17 years old. Students H and W were at the grade level of secondary 5, Students M and S were at secondary 1 and 3, respectively.

**NLG Tools**

We developed four NLG tools for the study so that students could try these tools and select the tools that most suit their idea generation strategies. We developed the tools on Hugging Face, a repository for open-source language models and machine learning applications. Students could freely access and repeatedly use the tools on Hugging Face. Table 1 provides details on the NLG tools. The variables by which each NLG tool differs are the number of language models in the tool, the sophistication of the tool's language model(s), the number of output texts, and the length of output text(s). We did not develop NLG tools using the largest and most sophisticated language models available in the marketplace, such as GPT-3 (Brown et al., 2020) and ChatGPT (OpenAI, 2022, November 30). Figures 3, 4 and 5 provide examples of the user interface of the NLG tools.

**Table 1**

*A summary of NLG tools used in this study*

| Tool name | Next Sentence Generator | Next Word Generator | Next Paragraph Generator 1 | Next Paragraph Generator 2 |
|---|---|---|---|---|
| Website | https://huggingface.co/spaces/Wootang01/next_sentence | https://huggingface.co/spaces/Wootang01/word_generator | https://huggingface.co/spaces/Wootang01/text_generator | https://huggingface.co/spaces/Wootang01/text_generator_two |
| SDK | Gradio | Streamlit | Gradio | Gradio |
| Language model(s) | GPT-J 6B (Biderman & Raff, | BERT base model (Devlin et al., 2019) | GPT-Neo 1.3B (Black et al., 2021) | GPT-Neo 2.7B |

| | 2022); GPT-Neo 2.7B (Black et al., 2021); GPT2-Large (Radford et al., 2019) | | | |
|---|---|---|---|---|
| Number of parameters | 6 billion; 2.7 billion; 774 million | 110 million | 1.3 billion | 2.7 billion |
| Maximum length of output | Unavailable | One word | 100 characters | 100 characters |
| Number of outputs | Three | One to seven | One | One |

**Figure 3**

*A screenshot of Next Sentence Generator*

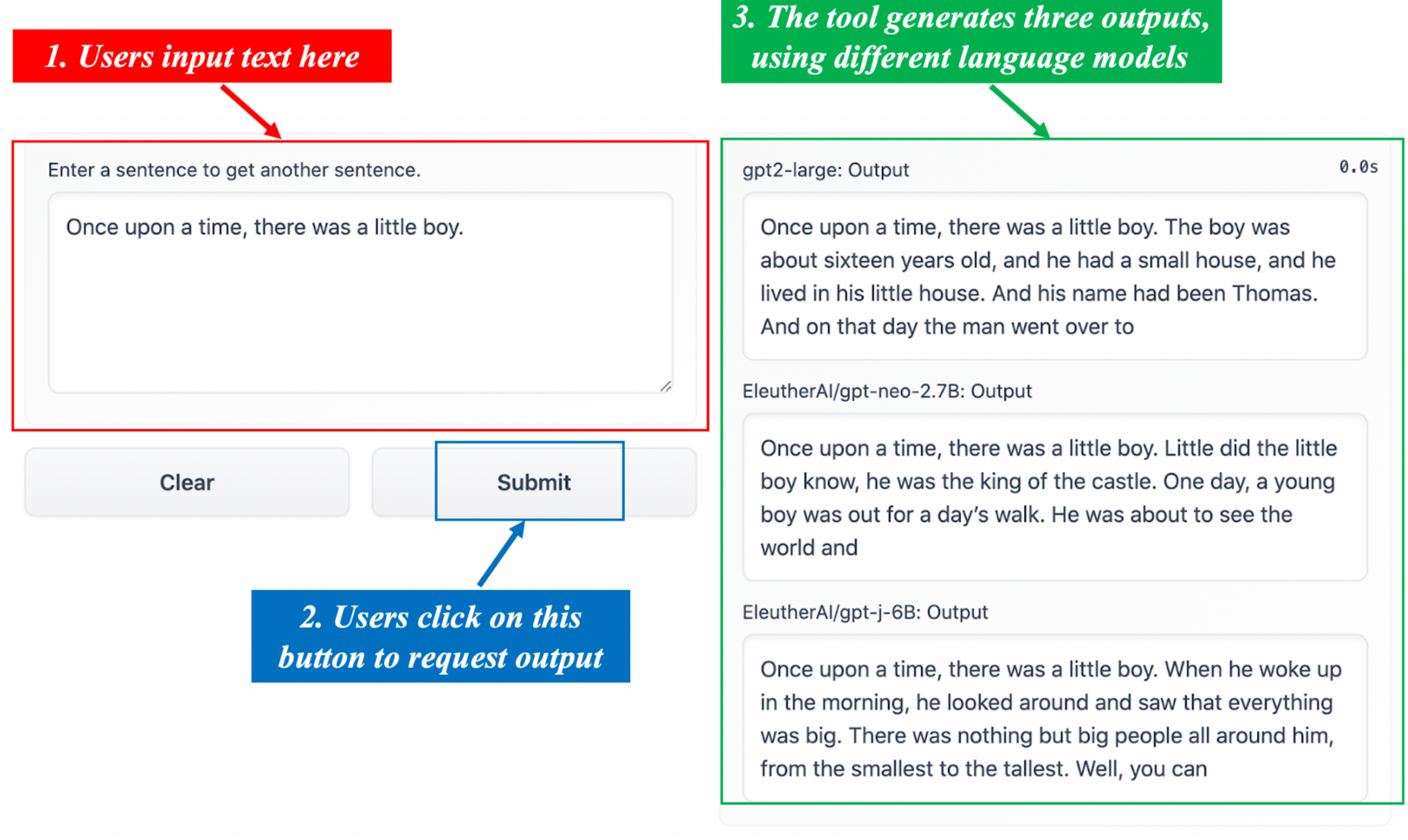

**Figure 4**

*A screenshot of Next Word Generator*

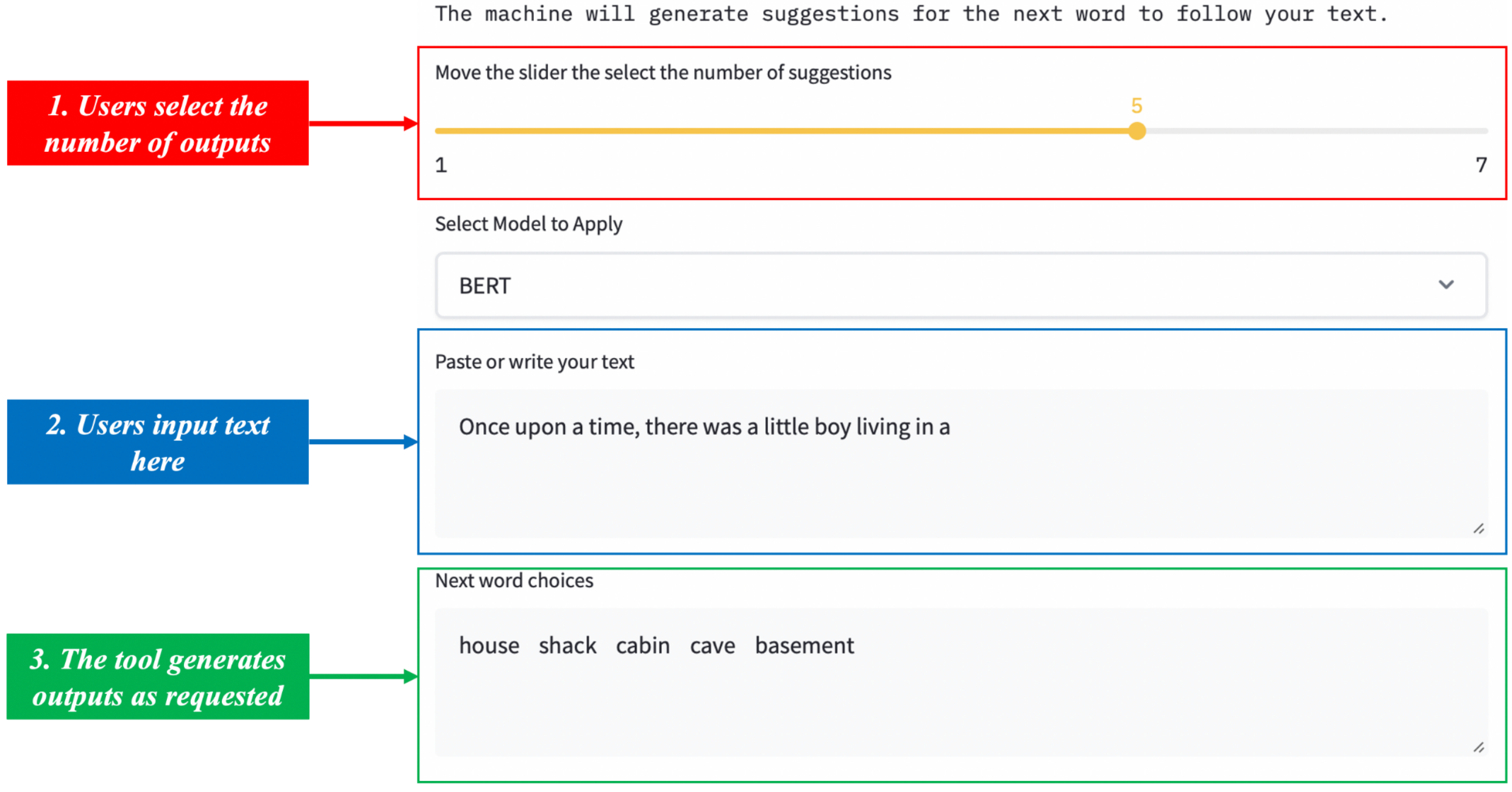

**Figure 5**

*A screenshot of Next Paragraph Generator 1*

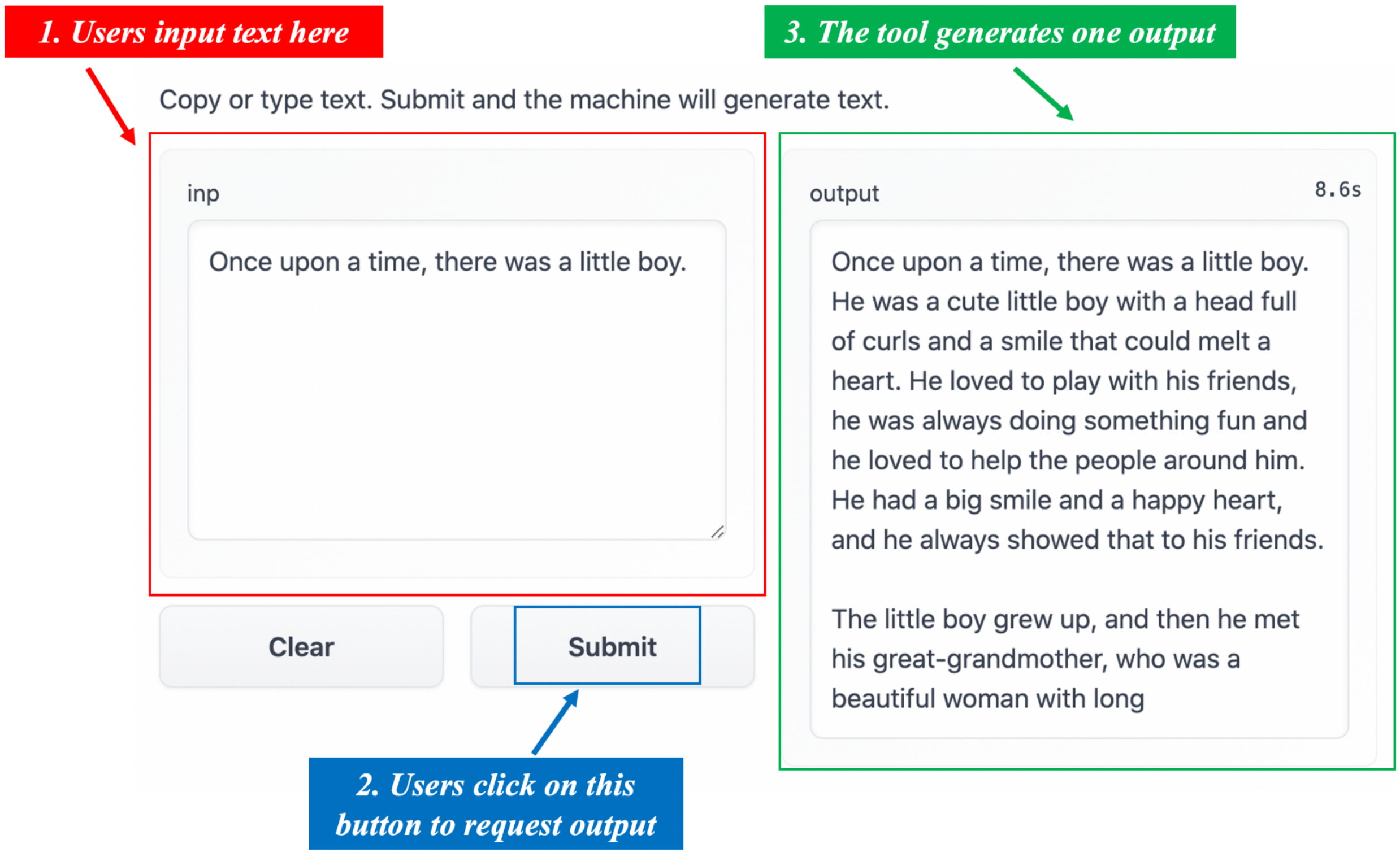

## Data Collection

Since we sought students' strategies from their perspective, we formulated six reflection questions for students to answer, aligning each reflection question with one or more of the research questions (See Table 2). By answering the questions, students could provide an introspective analysis of their creative writing with NLG tools. In addition, we asked students to provide screenshots so as to help us understand their written answers.

**Table 2**

*Workshop reflection questions*

| No. | Reflection question | Research Question No. | | |
|---|---|---|---|---|
| | | 1 | 2 | 3 |
| 1 | What was your plan today to write a short story using your own words and a machine's words? | ✓ | ✓ | ✓ |
| 2 | Please share a screenshot of your text generator input and output. How did you decide which words, sentences or paragraphs to put into a text generator? | ✓ | | |
| 3 | How do you feel writing a short story using your own words and a machine's words? | | ✓ | ✓ |
| 4 | If you could only use one text generator to write your short story, which would you use? And why? | | ✓ | ✓ |
| 5 | Please share a screenshot of your text generator input and output.<br>How did you decide which AI words, sentences or paragraphs to include in your story? | | ✓ | |
| 6 | Please share a screenshot of your text generator input and output. Why did you choose that text generator? | | | ✓ |

We sought students' strategies after each instance of their creative writing with NLG tools for which reason we had students answer reflection questions at the end of a workshop. Students were required to answer reflection questions on the same Google doc on which they wrote and submitted their short story. From the four participants, we collected 12 instances of students' answering reflection questions (three from Student H, four from Student M, two from Student S, and three from Student W), which were the main data source of this study. In addition, we collected the audio and video recordings of the workshops so as to triangulate students' written answers to reflection questions with what students shared verbally during workshop reflection time.

**Data Analysis**

We performed a thematic analysis (Braun & Clarke, 2006) on students' answers to the reflection questions. Our analysis aimed to uncover patterns within the students' answers, providing valuable insights into common themes in EFL students' idea generation during the task of story writing. To operationalize our analysis, we designed a coding scheme using an inductive approach, which enabled us to remain open to the data and to identify themes that were truly reflective of the students' answers, thereby enhancing the validity and relevance of our findings (Saldaña, 2012).

We aimed to capture elements of students' strategies when they 1) searched for ideas, 2) evaluated ideas and 3) selected NLG tools. First, we read through all answers to get a sense of the range of ideas. We then conducted open coding, generating codes and assigning those codes to relevant excerpts. For example, to identify elements of students' strategies for idea searching, we first looked for key words describing the content of prompts. To identify elements of students' strategies for idea evaluation, we looked for the key words describing how students read AI-generated ideas and what AI-generated ideas students copied. To identify elements of students' strategies to select NLG tools, we first looked for names of the NLG tools or any reference to an NLG tool in students' answers. When necessary, we referred to screenshots that students included in their reflections and audio and video recordings. In this coding process, the first author initial-coded and the second and third authors reviewed the coding scheme, discussing with the first author to resolve coding disparities, to propose new codes and to revise the coding scheme. After we agreed on the final codes, we compiled the codes in Online Resource 1.

After coding the data, we reread coded excerpts to help us gain a comprehensive understanding of the data and identify possible recurring patterns, themes, and ideas (Braun & Clarke, 2006). We looked for shared properties or relationships in the codes and elaborated themes that captured the essence of the coded data. After refining and redefining the themes, we arrived at clear and concise descriptions of each theme's scope, focus, and significance. We ensured that they accurately represented the coded data and explored how the themes were connected to each other and to the research questions (Braun & Clarke, 2006).

## Findings

Our thematic analysis revealed one main theme for students' searching for ideas, two for evaluation of ideas and one for selection of NLG tools. The four themes provide the best synthesis of the patterns found in students' answers. Since the themes represent an integration of patterns across many student answers, the themes are not mutually exclusive. Thus, some student answers reflect multiple themes. Likewise, quotes were chosen to be representative examples of each theme but the selected quotes may be attributed to more than one theme.

Based on the themes for the three research questions, in Figure 6 we present an enriched conceptual framework of idea generation between a student and NLG tools. Including the earlier proposed conceptual framework (see Figure 1). We elaborate these themes in the following sections. When reporting a student's direct written reflection, we provide the workshop number and the reflection question from which that written reflection was taken in brackets.

**Figure 6**

*Students' idea generation strategies for writing with NLG tools*

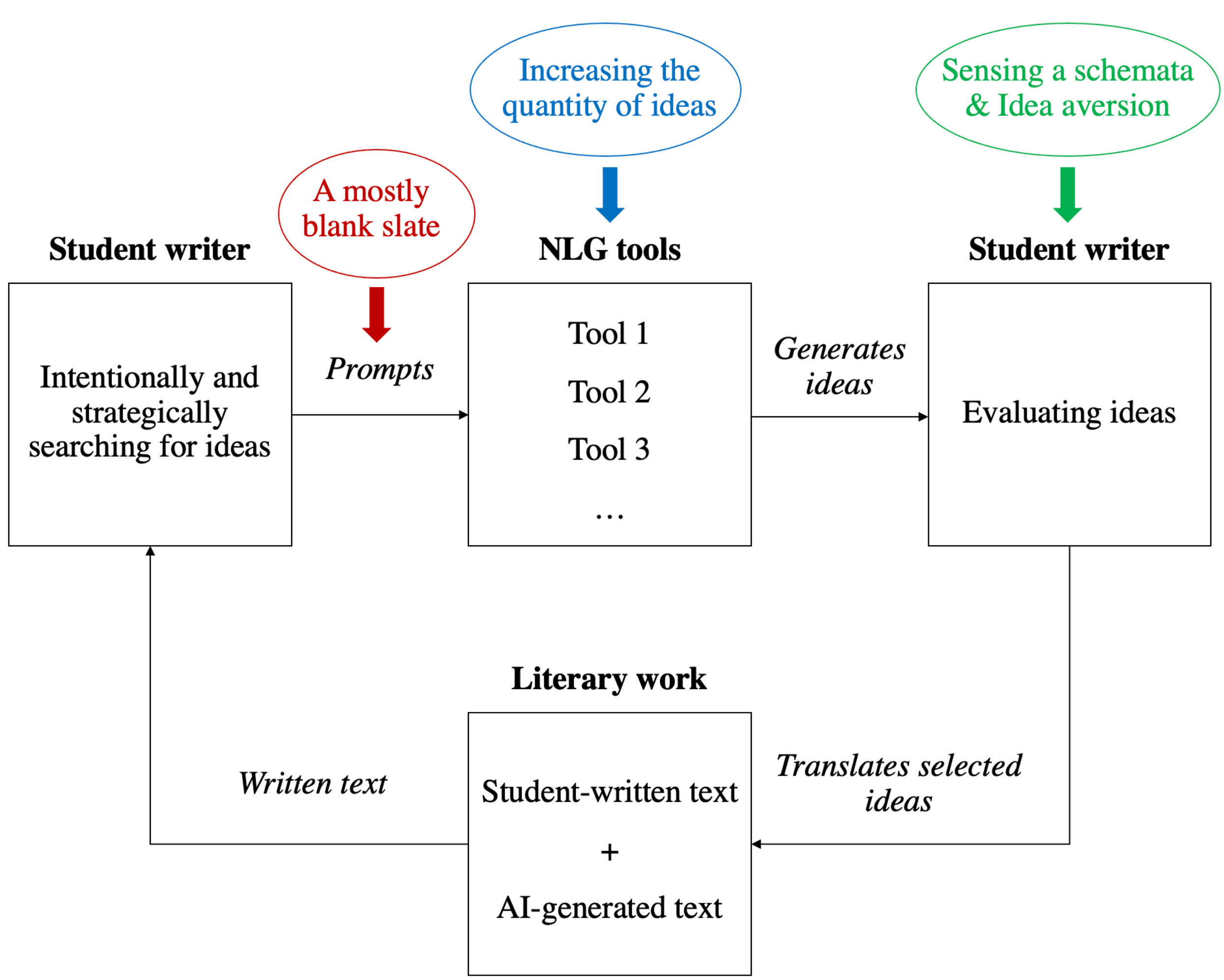

## Strategies For Searching for Ideas

### A mostly blank slate.

The first theme shows that although students may have had ideas when prompting an NLG tool, their search for story ideas appeared underdeveloped. This is because students described their search for ideas in terms of language features, but did not refer to discourse features, for instance, elements of a short story. For the reflection questions, students provided answers like, "When I can't think of the next word or sentence, I will copy the word or sentence then paste it in the input of the generator." (Student W, workshop three, reflection question two)

Moreover, students described the content of their prompts in terms of language features, not discourse features. On the one hand, Student M appeared to prompt NLG tools with the same level of language – a sentence. She wrote, "first I would think of a single sentence about the writing which is incomplete, then I would put it into the generator and let it do its work," (workshop three, reflection question two) and then, "I put a sentence at one time" (workshop seven, reflection question two). On the other hand, Students W and S prompted NLG tools with different levels of language. For instance, Student S wrote, "First, I put the whole *paragraph*, if the result is not what I look for, I cut it to the *last sentence*" (workshop three, reflection question two).

## Strategies For Evaluating Ideas

### Sensing a schemata.

A common approach to evaluating ideas from NLG tools was trying to place them in existing ideas or elements of a story. For example, Student S said, "I have to make sure I have the story in mind so that when I put in the text in the generator I can easily pick out a sentence that I want to use." (workshop two, reflection question one). And Student W said, "When the AI words or sentence matches my idea or the content of my story" (workshop three, reflection question five). And "I feel like it provided a lot of idea and also the topic sentence or the starting of the story" (workshop four, reflection question three).

#### Idea aversion.

This theme shows some students appeared more averse to novel or transformative ideas than other students. On the one hand, Student H said, "there were some words that I didn't want or expect but there were useful words to help me to make my piece more creative." (workshop two, reflection question three) and, "I would choose the sentence generator because with that I would be on the same page. The topic (output) would not be that off" (workshop two, reflection question four). Likewise, Student M said, "I would use the sentence generator since the words I will get in return will increase and the probability of words making sense inside the text may be bad or weird but it's still worth a try" (workshop three, reflection question four) and "the ones that make sense" (workshop seven, reflection question five). On the other hand, Student W said, "The next sentence generator, because it provides a wide variety of ideas." (workshop three, reflection question four).

### Strategies For Selecting NLG Tools

#### Increasing the quantity of ideas.

This theme shows students chose particular NLG tools because those tools produced a greater quantity of ideas than other tools. In this way, we found no one had preferred the Next Word Generator. Instead, Student S selected Next Paragraph Generators 1 and 2 because she could "pick out the best sentences to use" (workshop two, reflection question two) and "there's a few sentences I could use" (workshop three, reflection question one). Student M similarly said,

"I would use the sentence generator since the words I will get in return will increase" (workshop three, reflection question four).

## Discussion

### Major Findings

Collaboration with prompt-based NLG tools is an innovative approach to enhance EFL students' creative writing. This study investigated students' strategies of intentionally and strategically searching for ideas with NLG tools, of evaluating ideas generated by NLG tools and of selecting NLG tools. From a thematic analysis of students' written reflections, we found students may have existing ideas when searching for ideas with NLG tools but these ideas appear framed as language features not story elements. Second, we found students could evaluate ideas from NLG tools in terms of students' existing ideas or story elements; and students showed different degrees of aversion to novel or transformative ideas generated by NLG tools. Finally, we found students selected NLG tools that generated a greater number of ideas which students could evaluate.

### Implications

These findings are important theoretical contributions to understanding idea generation strategies for creative writing with NLG tools. This is because they shed light on the strategies of EFL students, who complete literary work not in their first language and who struggle with idea generation (Lin & Chang, 2020). Our study shows that EFL students have language concerns when searching for ideas with NLG tools and evaluating ideas from NLG tools; and they are concerned with the quantity of ideas generated by NLG tools. These concerns about language and the quantity of ideas may be unique to how EFL students approach creative writing with NLG tools. These concerns may not be shared by the educated, experienced English language writers in other creative writing studies with NLG tools (Calderwood et al., 2018; Yang et al., 2022).

Practically, the findings can inform creative writing instruction with prompt-based NLG tools for EFL students. Particularly, they can inform idea generation scaffolds that personalize learning (Zhai et al., 2021), that is, enable students who show different learning trajectories and different paces to learn with AI (Salas-Pilco, 2020). For example, at the point of searching for ideas with NLG tools, students who severely lack ideas might be instructed to prompt an NLG tool in any way for an immediate injection of ideas, be they words, sentences or even paragraphs; whereas students who have existing ideas might be instructed to consider these ideas in terms of story elements, such as characters, setting or events, and to search for specific ideas that logically fit their stories (Calderwood et al., 2018). At the point of evaluating ideas, students might benefit from receiving explicit instruction on making sense of the output in terms of existing ideas or a framework of story elements. Selecting story-specific ideas from AI-generated text resonates with existing findings on strategies used by highly educated adult writers (Calderwood et al., 2018; Yang et al., 2022). At the same time, although students' literary work should aspire to be art (Dawson, 2005), some students may show aversion to original or transformative ideas generated by NLG tools: these students might benefit from instruction to be open to unique ideas from NLG tools. As for selecting NLG tools for classroom writing, students might benefit from access to NLG tools that can produce lengthy texts such as ChatGPT or GPT-3 (Brown et al., 2020). These types of NLG tools might provide many ideas that writers can swap out or replace frequently (Calderwood et al., 2018) and that might increase writer productivity (Lee et al., 2022). The idea search and evaluation strategies that we recommend can also be applied to ChatGPT or GPT-3.

Finally, we recommend that whenever possible students plan ideas before interacting with an NLG tool. In other words, the more informed a student is about a story, the more the student can make informed decisions about searching for ideas and evaluating ideas from NLG tools. Likewise, NLG tools are not a panacea for students' idea generation difficulties but a scaffold, alongside writing plans, peer support and other scaffolds. Taken together, these scaffolds can inform design and implementation of the technology-supported learning activities that can improve students' creative writing (Shadiev et al., 2022), adding to the inspirational and instructional conditions that support creative writing and that lead to more creativity in students than without such conditions (Rahimi & Shute, 2021).

## Limitations and Future Directions

The study is exploratory in nature and its sample of four students is small so that its findings are not statistical generalizations but theoretical contributions to our understanding of EFL students' strategies in using NLG tools for creative writing. Thus, the themes found in this study are not exhaustive of all possible themes for idea generation. Subsequent research should test this study's contributions and attempt statistical generalizations, not least with a larger number of EFL students, a more diverse sample of students and more rigorous instrumentation. This larger, more diverse sample could account for a range of English language writing ability and digital literacy to use NLG tools. To recruit such students may require advertising in students' first language and not English language, or compulsory participation. Furthermore, as this study's findings are drawn largely from students' introspective analysis, subsequent research could instead capture data by screen recording students' interactions with NLG tools and audio recording students' narration during their interactions with NLG tools.

Another limitation is the technical features of the NLG tools. Although the open-source language models used in this study's NLG tools are state-of-the art, large language models are superior at present and we foresee advances in performance for open-source language models. Besides, the open-source language models in this study were neither fine-tuned nor were their default parameters changed.